\documentclass[iop,numberedappendix,twocolappendix,appendixfloats]{emulateapj}
\usepackage{natbib}
\usepackage{graphicx}
\usepackage{epsfig}
\usepackage{dcolumn}
\usepackage{bm}
\usepackage{epsf}
\usepackage{verbatim}
\usepackage{amsmath}
\usepackage{amssymb}
\usepackage{amsfonts}
\usepackage{cutwin}
\usepackage{xcolor}
\usepackage{fancyhdr}
\usepackage{wasysym}
\usepackage{mathtools}
\usepackage{float}
\usepackage{rotating}
\usepackage{hyperref}
\usepackage{etoolbox}

\newcommand{\be}{\begin{equation}}
\newcommand{\ee}{\end{equation}}
\newcommand{\msun}{{\rm M}_\odot}
\def \pth {}

\begin{document}

\submitted{}
\title{Flash Spectroscopy: Emission Lines from the Ionized Circumstellar Material around $<10$-Day-Old Type II Supernovae }

\author{
D.~Khazov\altaffilmark{1},
O.~Yaron\altaffilmark{1},
A.~Gal-Yam\altaffilmark{1},
I.~Manulis\altaffilmark{1},
A.~Rubin\altaffilmark{1},
S.~R.~Kulkarni\altaffilmark{2},
I.~Arcavi\altaffilmark{3,4},
M.~M.~Kasliwal\altaffilmark{2},
E.~O.~Ofek\altaffilmark{1},
Y.~Cao\altaffilmark{2},
D.~Perley\altaffilmark{2},
J.~Sollerman\altaffilmark{5},
A.~Horesh\altaffilmark{1},
M.~Sullivan\altaffilmark{6},
A.~V.~Filippenko\altaffilmark{7},
P.~E.~Nugent\altaffilmark{7,8},
D.~A.~Howell\altaffilmark{3,9},
S.~B.~Cenko\altaffilmark{10},
J.~M.~Silverman\altaffilmark{11},
H.~Ebeling\altaffilmark{12},
F.~Taddia\altaffilmark{5},
J.~Johansson\altaffilmark{5},
R.~R.~Laher\altaffilmark{13},
J.~Surace\altaffilmark{13},
U.~D.~Rebbapragada\altaffilmark{14},
P.~R.~Wozniak\altaffilmark{15},
T.~Matheson\altaffilmark{16}
}

\altaffiltext{1}{Benoziyo Center for Astrophysics, Faculty of Physics, The Weizmann Institute for Science, Rehovot 76100, Israel.}
\altaffiltext{2}{Astronomy Department, California Institute of Technology, 1200 E. California Boulevard, Pasadena, CA 91125, USA.}
\altaffiltext{3}{Las Cumbres Observatory Global Telescope Network, 6740 Cortona Drive, Suite 102, Goleta, CA 93111, USA}
\altaffiltext{4}{Kavli Institute for Theoretical Physics, University of California, Santa Barbara, CA 93106, USA.}
\altaffiltext{5}{The Oskar Klein Centre, Department of Astronomy, Stockholm University, AlbaNova, 10691 Stockholm, Sweden.}
\altaffiltext{6}{School of Physics and Astronomy, University of Southampton, Southampton, SO17 1BJ, UK.}
\altaffiltext{7}{Department of Astronomy, University of California, Berkeley, CA  94720-3411, USA.}
\altaffiltext{8}{Computational Research Division, Lawrence Berkeley National Laboratory, 1 Cyclotron Road MS 50B-4206, Berkeley, CA 94720, USA.}
\altaffiltext{9}{Department of Physics, University of California, Santa Barbara, CA 93106, USA.}
\altaffiltext{10}{Astrophysics Science Division, NASA Goddard Space Flight Center, Mail Code 661, Greenbelt, MD 20771, USA.}
\altaffiltext{11}{Department of Astronomy, University of Texas at Austin, Austin, TX 78712, USA.}
\altaffiltext{12}{Institute for Astronomy, University of Hawaii, 2680 Woodlawn Drive, Honolulu, HI 96822, USA.}
\altaffiltext{13}{Spitzer Science Center, California Institute of Technology, M/S 314-6, Pasadena, CA 91125, USA.}
\altaffiltext{14}{Jet Propulsion Laboratory, California Institute of Technology, USA.}
\altaffiltext{15}{Los Alamos National Laboratory, Los Alamos, New Mexico 87545, USA.}
\altaffiltext{16}{National Optical Astronomy Observatory, 950 North Cherry Avenue, Tucson, AZ 85719, USA}


\begin{abstract}

Supernovae (SNe) embedded in dense circumstellar material (CSM) may show prominent emission lines in their early-time spectra ($\leq 10$ days after the explosion), owing to recombination of the CSM ionized by the shock-breakout flash.
From such spectra (``flash spectroscopy''), we can measure various physical properties of the CSM, as well as the mass-loss rate of the progenitor during the year prior to its explosion.
Searching through the Palomar Transient Factory (PTF and iPTF) SN spectroscopy databases from 2009 through 2014, we found 12 Type II SNe showing flash-ionized (FI) signatures in their first spectra. All are younger than 10 days. These events constitute 14\% of all 84 SNe in our sample having a spectrum within 10 days from explosion, and 18\% of SNe~II observed at ages $<5$ days, thereby setting lower limits on the fraction of FI events.
We classified as ``blue/featureless" (BF) those events having a first spectrum which is similar to that of a black body, without any emission or absorption signatures. It is possible that some BF events had FI signatures at an earlier phase than observed, or that they lack dense CSM around the progenitor. Within 2 days after explosion, 8 out of 11 SNe in our sample are either BF events
or show FI signatures.
Interestingly, we found that 19 out of 21 SNe brighter than an absolute magnitude $M_R=-18.2$ belong to the FI or BF groups, and that all FI events peaked above $M_R=-17.6$ mag, significantly brighter than average SNe~II. 

\end{abstract}

\section{Introduction}

Core collapse (CC) in massive stars  ($M\gtrsim 8\,\msun$) results in various types of supernova (SN) explosions. Classification of such supernovae (SNe) is based primarily on their spectroscopic features (see, e.g., \citealt{1997ARA&A..35..309F} for a review). In general, Type II SNe show prominent lines of hydrogen in their spectra, and Type IIb also exhibit substantial He. Progenitors devoid of hydrogen result in Type Ib SNe which show helium lines, or Type Ic SNe which reveal neither hydrogen nor helium. SNe~Ic having very broad lines in their spectra [(2--3) $\times 10^4$\,km\,s$^{-1}$] are classified as Type Ic-BL (and some of these are associated with gamma-ray bursts).
Further classification among SNe~II is based on their visual-wavelength light curves. Type II-P SNe show a plateau of $\sim 100$ days duration, while Type II-L show a more linear (in magnitudes) decay. 
 The distinction between Types II-L and II-P is somewhat controversial \citep{2012ApJ...756L..30A, 2014ApJ...786...67A, 2015ApJ...799..208S, 2014MNRAS.445..554F}.

Along their evolution and particularly during the post-main-sequence lifetime, massive stars tend to lose mass either by winds, binary interactions, or various eruptive events \citep[e.g.,][]{2014ARA&A..52..487S}.
If the mass is lost from the progenitor shortly before it explodes as a SN, the interaction between the SN ejecta and the circumstellar material (CSM) may result in X-ray or radio emission for normal winds, and in narrow optical emission lines for a dense CSM. The latter are classified as Type IIn (H lines; e.g., \citealt{1990MNRAS.244..269S, filippenko91, 2012ApJ...744...10K, 2013A&A...555A..10T}) or Type Ibn (He lines; e.g., \citealt{2008MNRAS.389..113P}) SNe.

Intense mass loss via eruptive events sometimes manifests itself as precursors prior to the SN explosion \citep{2014ApJ...789..104O}, as in SN 2006jc \citep{2007Natur.447..829P}, SN 2010mc \citep{2013Natur.494...65O}, PTF11qcj \citep{2014ApJ...782...42C}, and possibly SN 2009ip \citep[whether a SN was finally produced is controversial; e.g.,][]{2013MNRAS.430.1801M, 2014ApJ...789..104O, 2014ApJ...787..163G, 2014ApJ...780...21M, 2015AJ....149....9M}.
Mass loss from massive stars has an important role in their evolution, and is critical in determining the type of the SN explosion. 
The physics and rates of mass loss are generally not well understood, and the rates used in stellar evolution models are quite uncertain \citep{2012ARA&A..50..107L, 2014ARA&A..52..487S}.

Optical spectra obtained a short time (a few days or less) after a SN explosion may be dominated by features generated by ionization of the CSM by ultraviolet (UV) radiation emitted during the hot shock breakout and the early shock-cooling phase \citep[SN 2013cu = iPTF13ast;][]{2014Natur.509..471G}. The photons emitted from the SN ionize the material around the progenitor (if such exists), and this CSM recombines and radiates strong emission lines. It is possible that further emission is generated by collisional excitations by the free electrons. From a series of spectra showing these emission lines, we can learn about the elemental abundances in the CSM, the early temperature evolution of the ejecta, and the progenitor mass-loss rate shortly prior to the explosion \citep[e.g.,][]{2014A&A...572L..11G, 2015ApJ...806..213S, 2015MNRAS.449.1876S}. 
We refer to the method of obtaining such early-time spectra as ``flash spectroscopy'' and the spectra showing such emission features as ``flash ionized'' (FI) \citep{2014Natur.509..471G}.
Previous examples of early-time spectra that are flash ionized were presented for SN 1983K \citep{1985ApJ...289...52N, 1990PASP..102..299P}, SN 1993J \citep{1994AJ....108.1002G, 2000AJ....120.1487M}, SN 1998S \citep{2000ApJ...536..239L, 2015ApJ...806..213S}, and SN 2006bp \citep{2007ApJ...666.1093Q}.

Following the detection of flash-ionized (FI) signatures in iPTF13ast \citep{2014Natur.509..471G}, we have searched for additional similar events in the Palomar Transient Factory \citep[PTF;][]{2009PASP..121.1334R, 2009PASP..121.1395L} and intermediate PTF (iPTF) spectroscopic databases from 2009 through the end of 2014, showing prominent high-ionization (in particular,  He~II $\lambda 4686$) emission lines in early-time spectra which later disappear. Our main motivation is to quantify the frequency of such objects, and hence the fraction of progenitors embedded in CSM.
We find that a substantial fraction of SNe show FI signatures in their early-time spectra, and the majority of SNe~II observed a few days after explosion have spectra that are either blue and featureless or exhibit prominent emission lines.

In Section \ref{sec:observations} we present the observations, the sample construction, and the SN classification criteria. Our results are given in Section \ref{sec:results}. In Sections \ref{sec:discussion} and \ref{sec:summary} we discuss and summarize the main ideas.

\section{Observations and Sample Construction}
\label{sec:observations}

The SN discoveries, classifications, and redshifts were obtained as part of the PTF and iPTF surveys. The SN discovery is confirmed by on-duty astronomers, who also trigger follow-up observations. A review of the real-time alert system and follow-up programs can be found in \cite{2011ApJ...736..159G}, and the latest machine-learning procedures in use (distinguishing a real source from a bogus artifact) are described in \citet{2013MNRAS.435.1047B, 2013AAS...22143105W} and \citet{2015AAS...22543402R}. UT dates are used throughout this paper.
All epochs are given in the observer's frame; the redshifts of the SNe in our sample are low ($z<0.2$), and the corrections are negligible for our purposes.
Table \ref{tab-sne} summarizes the details of the SNe whose spectra are presented in this paper: the 12 flash-ionized SNe and the 3 events younger than 2 days which are neither flash-ionized nor blue/featureless (see \S \ref{subsec: sampleConstruction}).

\subsection{Spectra}

All spectra used for this work were obtained as part of the PTF survey and were available via our internal (Marshal) database.
The details of the spectra are summarized in Table \ref{tab-spec}. 
These data are made public via the WISeREP portal \citep{2012PASP..124..668Y}. Spectral reductions were carried out using standard procedures. Full discussions of our spectroscopic datasets will be reported in future publications.

\begin{sidewaystable}
\tiny

\caption{15 PTF and iPTF Supernovae with Early-Time Spectroscopy}
\begin{tabular}{lll p{1.6cm} l p{1.7cm} p{1.3cm} p{1.5cm} p{1.5cm} p{1.5cm} p{1.7cm} p{1.8cm}}

Name & $\alpha$ [deg] & $\delta$ [deg] & Spectroscopic Classification & Redshift & First Detection Date\tablenotemark{a} & $R$ Detection Mag & Latest Nondetection Limit Date\tablenotemark{a} & Limiting Mag & Estimated Explosion Time & First Spectrum Date\tablenotemark{a} & References\\
\hline\hline

PTF09ij & 218.061018 & 54.855424 & SN II & 0.124 & 2009-05-20.29 [2.01] & 21.05 & 2009-05-16.28 [-2.01] & 21.82 & 2009-05-18.28 & 2009-05-21.28 [3] & \\
PTF10abyy & 79.16885 & 6.798268 & SN II & 0.0297 & 2010-12-03.32 [0.51] & 21.14 & 2010-12-02.31 [-0.51] & 21.88 & 2010-12-02.82 & 2010-12-09.6 [6.79] & \\
PTF10gva & 185.98082 & 10.580728 & SN II & 0.0276 & 2010-05-05.17 [0.9] & 18.13 & 2010-05-03.38 [-0.9] & 21 & 2010-05-04.27 & 2010-05-06.4 [2.12] & \\
PTF10gvf & 168.438496 & 53.629126 & SN IIn & 0.081 & 2010-05-04.31 [0.65] & 21.44 & 2010-05-03.66 [0] & N.A \tablenotemark{b} & 2010-05-03.66 & 2010-05-06.52 [2.86] & \cite{2014ApJ...789..104O}\\
PTF10tel & 260.377817 & 48.129834 & SN IIn & 0.035 & 2010-8-20.22 [2.5] & 21.99 & 2010-8-17.72 [0] & N.A \tablenotemark{b} & 2010-8-17.72 & 2010-08-26.26 [8.53] & SN 2010mc; \cite{2013Natur.494...65O}\\
PTF10uls & 20.344412 & 4.891319 & SN II & 0.0479 & 2010-09-07.43 [0.48] & 21.21 & 2010-09-06.48 [-0.48] & 21.78 & 2010-09-06.96 & 2010-09-10.44 [3.48] & \cite{2014ApJ...789..104O}\\
PTF11iqb & 8.52015 & -9.704979 & SN II & 0.0125 & 2011-07-23.37 [1.1] & 16.89 & 2011-07-22.27 [0] & N.A \tablenotemark{b} & 2011-07-22.27 & 2011-07-24.33 [2.06] & \cite{2015MNRAS.449.1876S, 2014ApJ...789..104O}\\
PTF12gnn & 239.705326 & 36.169707 & SN II & 0.0308 & 2012-07-09.36 [0.97] & 18.83 & 2012-07-07.43 [-0.97] & 20.61 & 2012-07-08.39 & 2012-07-12.36 [3.97] & \\
PTF12krf & 342.069469 & 24.149513 & SN II & 0.0625 & 2012-11-04.13 [0.99] & 20.65 & 2012-11-02.15 [-0.99] & 21.27 & 2012-11-03.14 & 2012-11-07.19 [4.05] & \\
iPTF13ast & 218.495242 & 40.239672 & SN IIb & 0.0258 & 2013-05-03.18 [0.16] & 20.22 & 2013-05-03.02 [0] & N.A \tablenotemark{b} & 2013-05-03.02 & 2013-05-03.5\tablenotemark{c} [0.48] & SN 2013cu; \cite{2014Natur.509..471G}\\
iPTF13dqy & 349.936251 & 10.184555 & SN II & 0.011855 & 2013-10-06.25 [0.45] & 18.63 & 2013-10-05.34 [-0.45] & 20.71 & 2013-10-05.79 & 2013-10-06.38 [0.59] & Yaron et al. (2015; Nature Physics, submitted)\\
iPTF14bag & 185.292993 & 64.343544 & SN II & 0.116 & 2014-05-18.23 [0.02] & 20.64 & 2014-05-18.19 [-0.02] & 21.37 & 2014-05-18.21 & 2014-05-21.34 [3.13] & \\
\hline
iPTF13aaz & 169.737245 & 13.063896 & SN IIP & 0.002692 & 2013-03-22.17 [1.42] & 16.26 & 2013-03-19.34 [-1.42] & 20.96 & 2013-03-20.75 & 2013-03-22.5 [1.74] & SN 2013am \cite{2014ApJ...797....5Z}\\
iPTF13dkk & 355.396485 & 3.725103 & SN II & 0.0092 & 2013-09-12.19 [0.35] & 18.91 & 2013-09-11.49 [-0.35] & 21.39 & 2013-09-11.84 & 2013-09-12.99 [1.15] & \\
iPTF14ayo & 181.512516 & 47.492568 & SN II & 0.0023 & 2014-05-14.16 [0.49] & 18.72 & 2014-05-13.19 [-0.49] & 20.7 & 2014-05-13.68 & 2014-05-14.5 [0.82] & \\

\end{tabular}
\footnotetext[1]{All dates are UT. In brackets: the time in days from the estimated explosion.}
\footnotetext[2]{We obtained the limit by fitting the rise time.}
\footnotetext[3]{This spectrum does not contain observed wavelengths below 4900\AA. We present a spectrum obtained a few days later.}

\label{tab-sne}
\end{sidewaystable}

\begin{sidewaystable}
\tiny

\caption{Log of Spectroscopic Observations}
\begin{tabular}{llllll p{3cm} l }

Name & UT Date & Telescope & Instrument & Observer & Reducer & He~II $\lambda$4686 EW (\AA) & Reference \\
\hline\hline
PTF09ij & 2009-05-21.28 & Palomar 5.1\,m & DBSP & Kasliwal & Kasliwal & $-6.85 \pm 0.48$ \\
PTF10abyy & 2010-12-09.60 & Keck I 10\,m & LRIS & Ebeling & Cenko & $-8.59 \pm 0.15$ \\
PTF10gva & 2010-05-06.39 & Keck I 10\,m & LRIS & Cenko et al. & Cenko & $-2.65 \pm 0.06$ \\
PTF10gvf & 2010-05-06.52 & Keck I 10\,m & LRIS & Cenko et al. & Cenko & $-4.83 \pm 0.34$ \\
PTF10tel & 2010-08-26.26 & Gemini N 8\,m & GMOS & Service & Howell, Murray & $-2.93 \pm 0.27$ \\
PTF10uls & 2010-09-10.44 & Kitt Peak 4\,m & RC Spec & Kulkarni, Cenko & Cenko & $-4.34 \pm 0.36$ \\
PTF11iqb & 2011-07-24.33 & Gemini S 8\,m & GMOS & Service & Parrent & -$7.21 \pm 0.18$ \\
PTF12gnn & 2012-07-12.36 & Lick 3\,m & Kast & Clubb, Filippenko & Silverman & $-3.37 \pm 0.12$ \\
PTF12krf & 2012-11-07.19 & Palomar 5.1\,m & DBSP & Horesh, Tang & Yaron & $-2.74 \pm 0.15$ \\
iPTF13ast & 2013-05-06.01 & Nordic Optical Tel. & ALFOSC & Geier (Service\tablenotemark{a}) & Taddia  & -4.58$\pm$0.13 \\
iPTF13dqy & 2013-10-06.38 & Keck I 10\,m & LRIS & Perley & Perley & $-10.30\pm 0.22$ \\
iPTF14bag & 2014-05-21.33 & Apache Point 3.5\,m & DIS & Cao & Cao & $-8.74\pm 0.41$ \\
\hline
iPTF13aaz & 2013-03-22.50 & Lick 3\,m & Kast & Fox et al. & Silverman & \\
iPTF13dkk & 2013-09-12.99 & Nordic Optical Tel. & ALFOSC & Taddia, Van Eylen & Johansson & \\
iPTF14ayo & 2014-05-14.29 & Gemini N 8\,m & GMOS & Service & Cenko & \\
\hline
SN 1993J & 1993-03-30 & Lick 3\,m & Kast & Davis, Schlegel & & -0.24 $\pm$ 0.05 & \citet{2000AJ....120.1487M} \\
SN 1998S & 1998-03-04 & Keck II 10\,m & HIRES & Matheson, Filippenko & Shivvers & -10.50 $\pm$ 0.10 & \citet{2015ApJ...806..213S}\\
SN 2006bp & 2006-04-11 & HET 9.2\,m & LRS & Quimby & & $-1.00 \pm 0.05$ & \citet{2007ApJ...666.1093Q}

\end{tabular}
\footnotetext[1]{ToO program, PI G. Leloudas.}
\label{tab-spec}
\end{sidewaystable}

\subsection{Photometry}

PTF survey images are obtained by the survey camera \citep{2008SPIE.7014E..4YR} mounted on the Palomar 48-inch Schmidt telescope in the Mould $R$ or Gunn $g$ bands. Transient discovery is performed by an automated real-time pipeline using image subtraction \citep[e.g.,][]{2011ApJ...736..159G}. The system provides preliminary automated photometry\footnote{This is internally referred to as the ``Nugent pipeline."} which can often be improved. Full processing is then conducted using the PTF/IPAC pipeline \citep{2014PASP..126..674L} and the data are photometrically calibrated \citep{2012PASP..124...62O}.

In this paper, we present photometry which has been processed through our custom image-subtraction pipelines\footnote{These are internally referred to as the ``Sullivan pipeline'' and PTFIDE.}, following the IPAC/PTF photometric pipeline. Our image-subtraction pipeline has been used extensively in earlier PTF science publications \citep[e.g.,][and references therein]{2015MNRAS.446.3895F}. The coadd reference for each object was constructed from images taken up to 20 days before the discovery, using the same filter as when the object was detected by the automated pipeline.
The pipeline subtracts the reference from every image, performs point-spread function (PSF) photometry on the result, and provides the calibrated flux and the flux uncertainty for each of the images, using the Sloan Digital Sky Survey (SDSS; when available) or PTF zero points. The detection threshold is set to be 3 times the $1\sigma$ flux uncertainty. All values have been converted to the native PTF photometric system \citep{2012PASP..124...62O}.

The photometry presented here is corrected for Galactic extinction by calculating $E_{B-V}$ according to \citet{1998ApJ...500..525S}, and the extinction following \citet{1989ApJ...345..245C}.
No K-corrections were applied, given the low redshifts of the SNe in our sample ($z<0.2$).

\subsection{Sample Construction and SN Classification}
\label{subsec: sampleConstruction}

\begin{figure*}[h]
\centering

\includegraphics[width=0.4\textwidth]{\pth 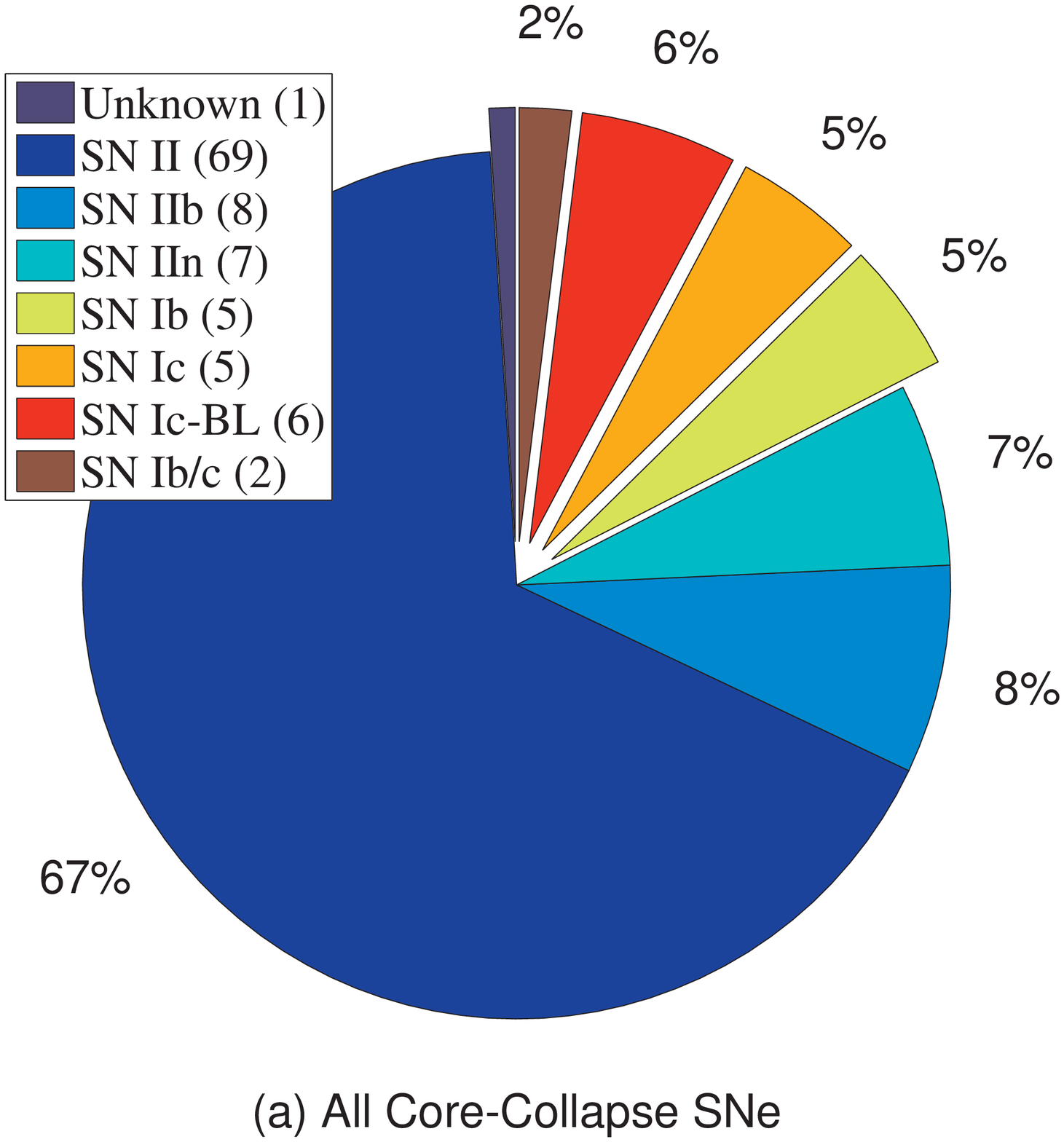}
\includegraphics[width=0.4\textwidth]{\pth 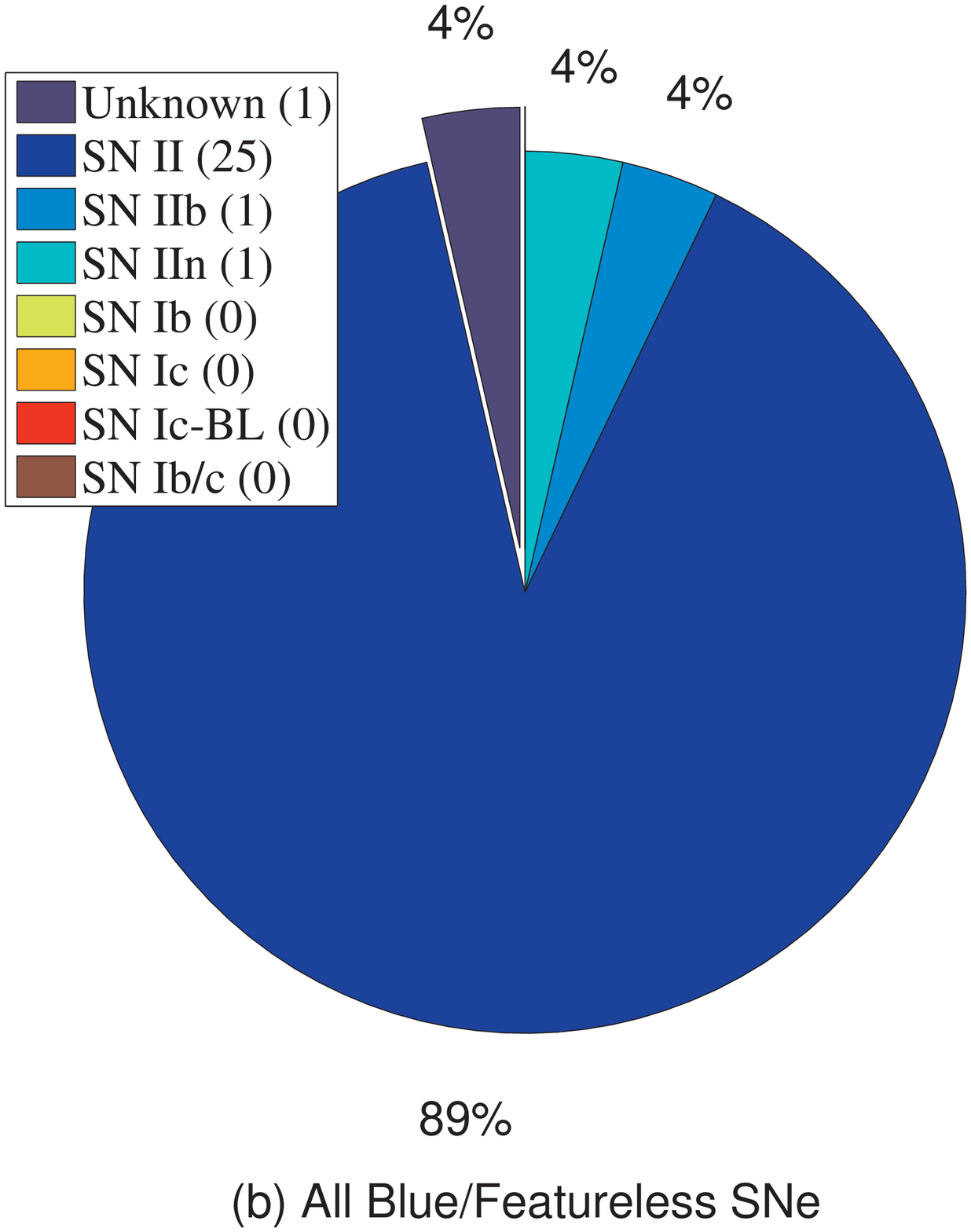}
%
%
\includegraphics[width=0.4\textwidth]{\pth 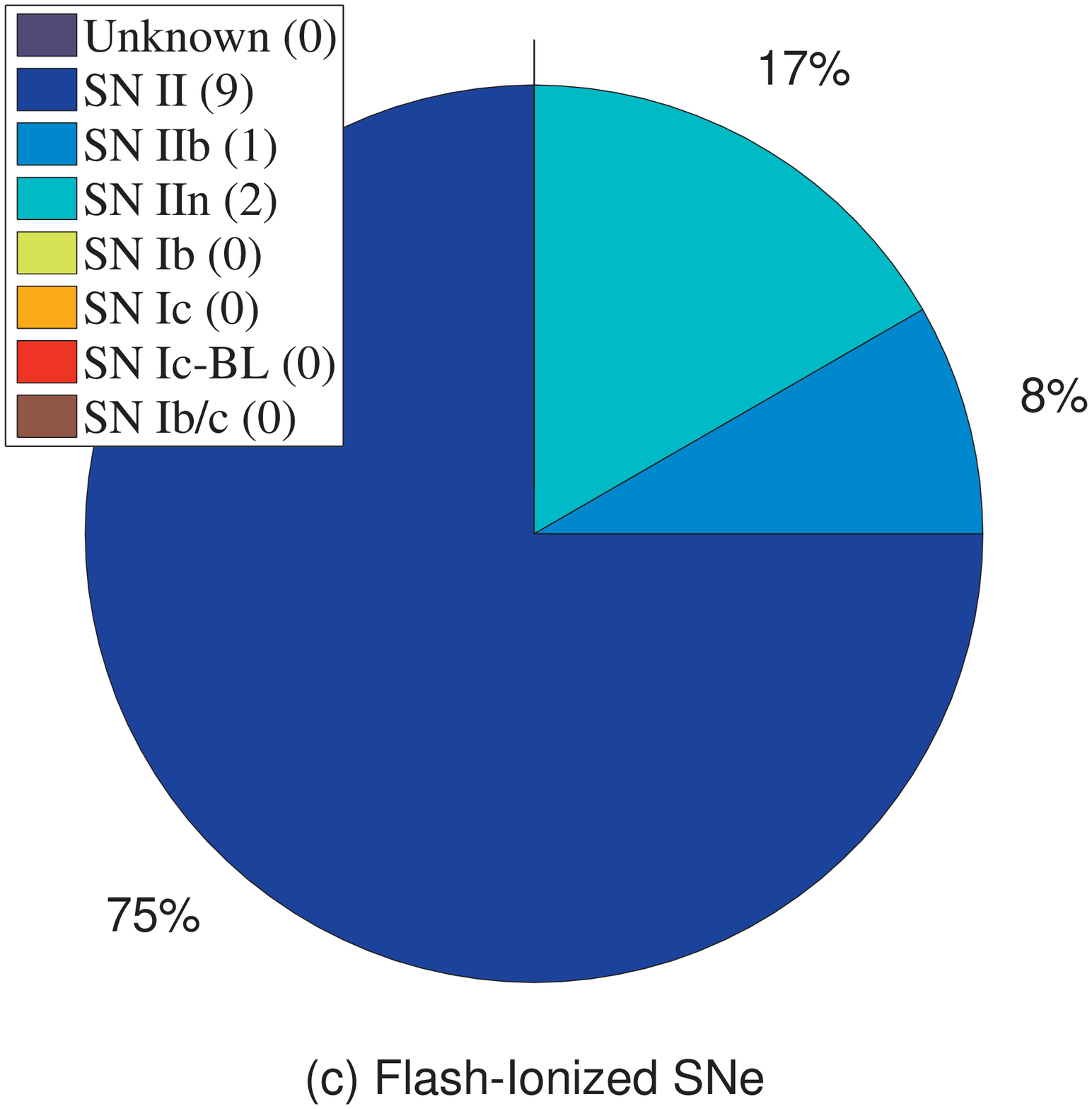}

\caption{(a) Distribution by type of all 103 CC SNe whose first spectrum was taken $<10$ days after the time of the pre-explosion limit. Of these, 84 are SNe~II. (b) Type distribution of events with BF first spectra (28 total). (c) Type distribution of FI SNe (12 total).
There is only a single, featureless spectrum of the ``unknown'' SN, not enough for classification.} 
\label{fig:pie}
\end{figure*}

Our sample consists of 103 CC SNe, 84 of them of Type II (Figure \ref{fig:pie}a), discovered by the PTF and iPTF surveys between 2009 and 2014, whose first spectra were obtained within 10 days from the time of the SN pre-explosion limit.
We chose the pre-explosion limit as the last nondetection where the limiting magnitude (the nondetection threshold) is fainter by 0.5\,mag than the first detection. 
For SNe with a well-sampled rise to peak, we used a parabolic fit and estimated the explosion time and the pre-explosion limit as the time when the flux is 0.
We decided to focus on SNe younger than 10 days after an initial inspection of the first spectra of SNe obtained within 6 days from the discovery. These events include a large population of old SNe (sometimes discovered a long time after their last nondetection), and we found that all of the events of interest (showing flash-ionized signatures) were younger than 10 days from the time of the pre-explosion limit.
For additional details see Appendix \ref{appen:sampleConstruction}.


For all of our events we use the improved photometry to estimate the explosion time, either using a parabolic fit as explained above, or as the mean between the times of the pre-explosion limit and the first detection.

\begin{figure*}[h]
\centering
\includegraphics[width=1\textwidth]{\pth 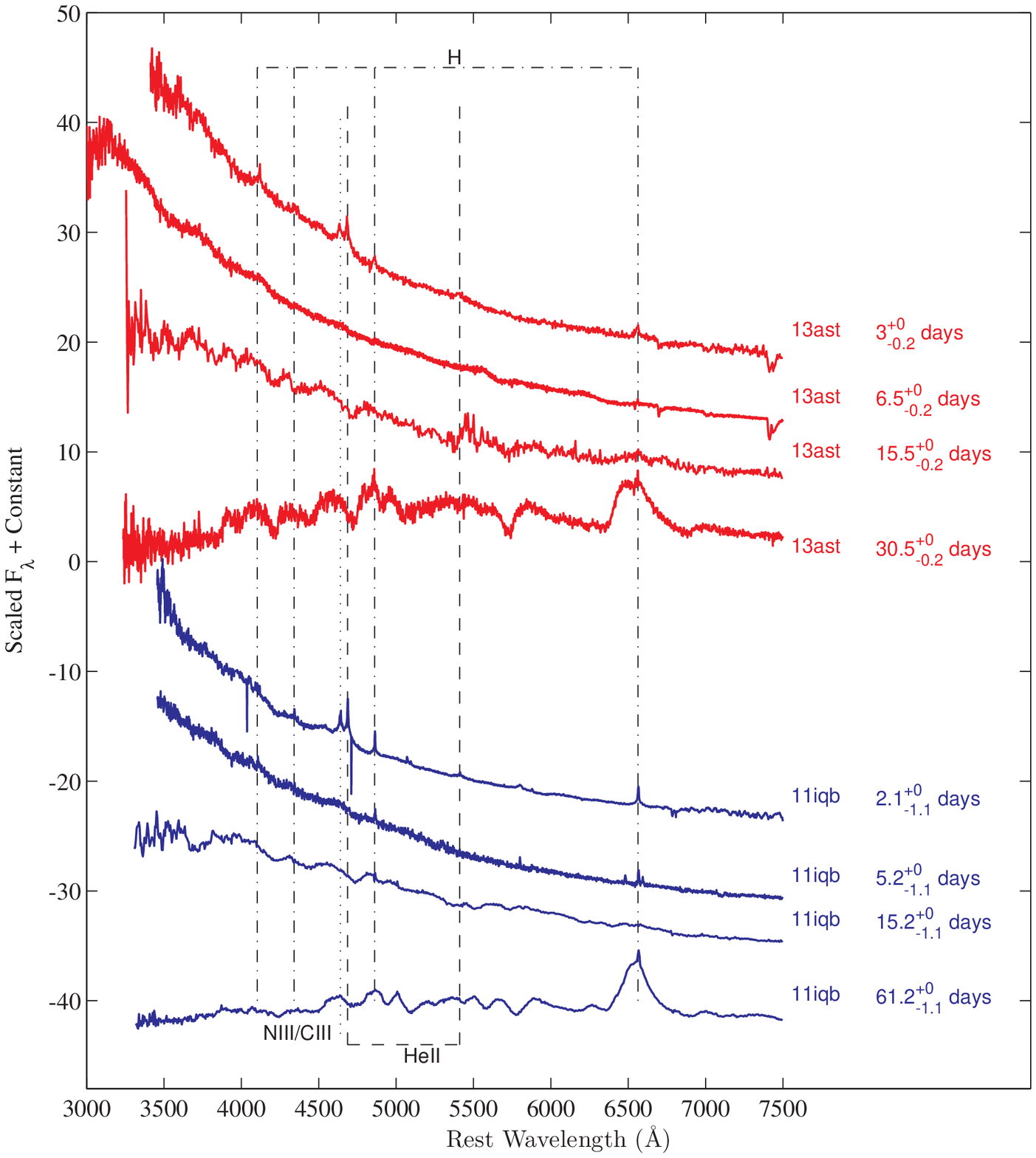}
\caption{Spectral evolution of two of the FI events. The emission lines disappear within a few days, and P-Cygni line profiles develop over time. These well-observed objects would be classified as FI if found prior to day 3 and as BF if found around day 5, demonstrating the evolution of SNe~II, and why our measured FI fraction is only a lower limit on the true frequency of SNe having dense CSM. Data from \citet{2014Natur.509..471G} and \citet{2015MNRAS.449.1876S}.}
\label{fig:flashSpecEvolution}
\end{figure*}

We then classified all SNe based on their earliest spectra as either ``blue/featureless'' (BF), ``flash ionized" (FI), or neither.
A BF spectrum resembles the spectrum of a black body, without any prominent emission or absorption lines, and without broad P-Cygni profiles  (ignoring host-galaxy emission lines, night-sky lines, or any features caused by imperfect spectral reduction; Figure \ref{fig:flashSpecEvolution}).
The spectra of SNe classified as FI show a blue continuum with various emission lines, including in particular He~II $\lambda 4686$, accompanied by H Balmer lines, with widths corresponding to a velocity dispersion of $\sim 1000$\,km\,s$^{-1}$.\footnote{The line is broadened by a combination of Doppler velocity and electron scattering which may be dominant (see, e.g., \citealt{2014A&A...572L..11G}).} Since the H lines can be easily contaminated by narrow host-galaxy emission but the He~II generally cannot, we focus on the He~II $\lambda$4686 line. Moreover, its width corresponds to velocities much lower than those associated with the SN explosion, so its presence provides good evidence for emission from CSM ionized by UV radiation.
We considered objects with persistent emission lines (SNe~Ibn or IIn) as having flash-ionization signatures only if certain features (in particular, the He~II line) disappeared in later spectra (Figure \ref{fig:flashSpecEvolution}).
We found that all events classified as FI or BF are Type II SNe\footnote{We exclude a peculiar object of uncertain nature that possibly shows FI signatures and is the subject of a forthcoming publication (Kasliwal et al., in prep.).}, so from now on we focus our discussion on the 84 SNe~II in our sample.


We quantify the level of He~II $\lambda 4686$ emission in spectra of the 84 objects included in our final sample by calculating the equivalent width (EW) of an emission line located in the range 4666--4706\,\AA. The continuum is fitted by a linear function in the range of 4545--4620\,\AA\ on one side and 4726--4800\,\AA\ on the other. In order to estimate the significance of any detection we conduct the following simulation. We randomly choose a continuous subrange of at least 30\% of our continuum area (yielding subranges with a mean of $\sim 40$\,\AA\ and a standard deviation of $\sim 12$\,\AA\ on each side) and recalculate the EW. By repeating this procedure 500 times, we estimate the uncertainty in the EW as the standard deviation of the results. We adopt as significant only cases where the relative error $\Delta$EW/EW $<0.1$ (attributing any weaker detections to noise or to uncertain continuum estimation). We classify as FI events those having EW $<0$. We list the He~II $\lambda 4686$ EW values in FI spectra in Table \ref{tab-spec}. We also calculated the EW values of He~II in the early-time spectra of previous events (except for SN 1983K, whose spectra are not available in electronic format), and they are presented as well in Table \ref{tab-spec}. These events would indeed be counted as FI events according to our selection criteria.

More details regarding the sample construction and the estimation of SN explosion times are given in Appendix \ref{appen:sampleConstruction}.

\section{Results}
\label{sec:results}


\subsection{Spectral Classification}

\begin{figure*}[h]
\centering
\includegraphics[width=1\textwidth]{\pth 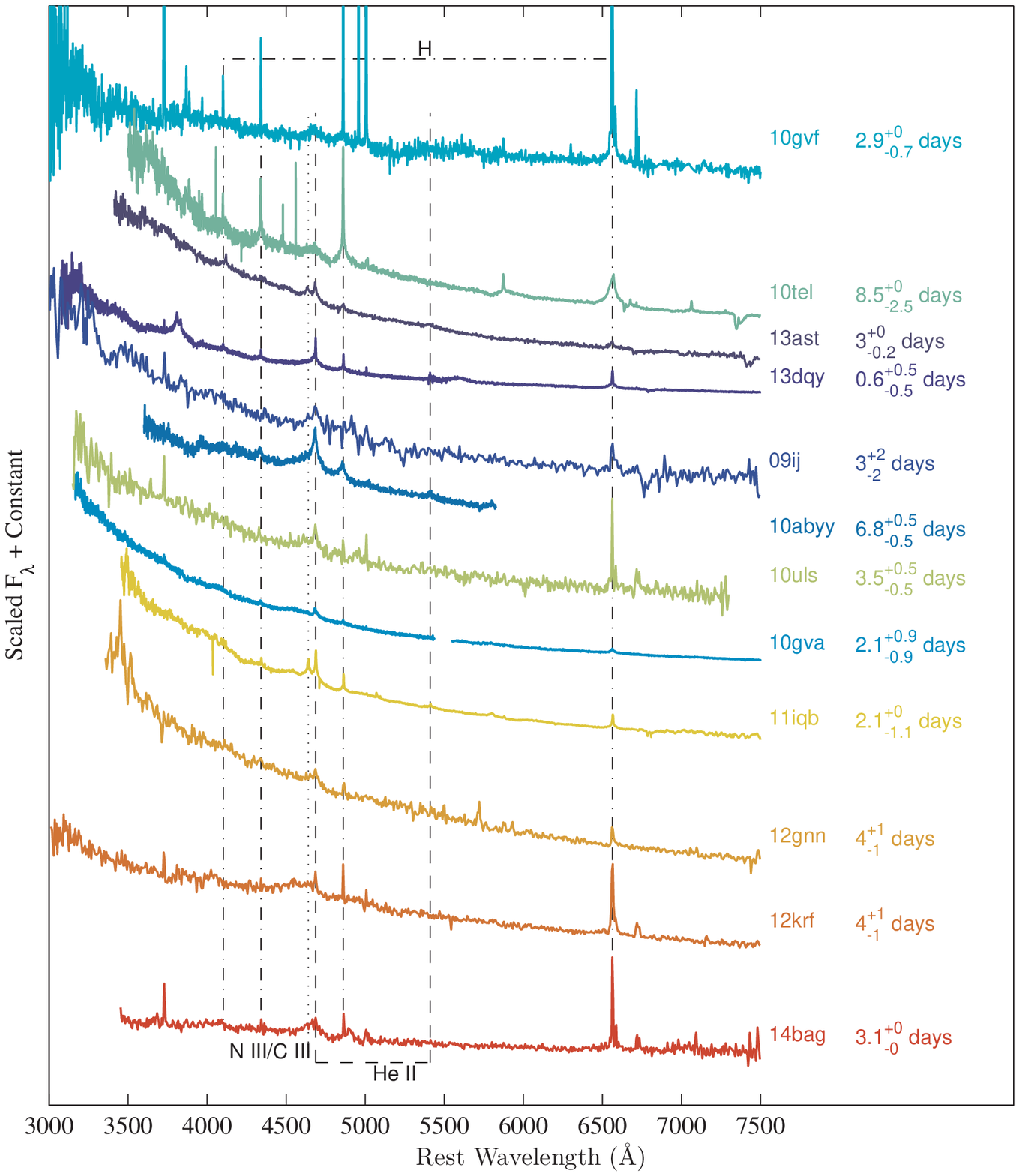}
\caption{Spectra of our twelve FI events. On the right: an estimate of the age of the SN, with respect to the estimated explosion time (see Appendix \ref{appen:sampleConstruction} for details).}
\label{fig:flashSpec}
\end{figure*}

\begin{figure*}[h]
\centering

\includegraphics[width=0.285\textwidth]{\pth 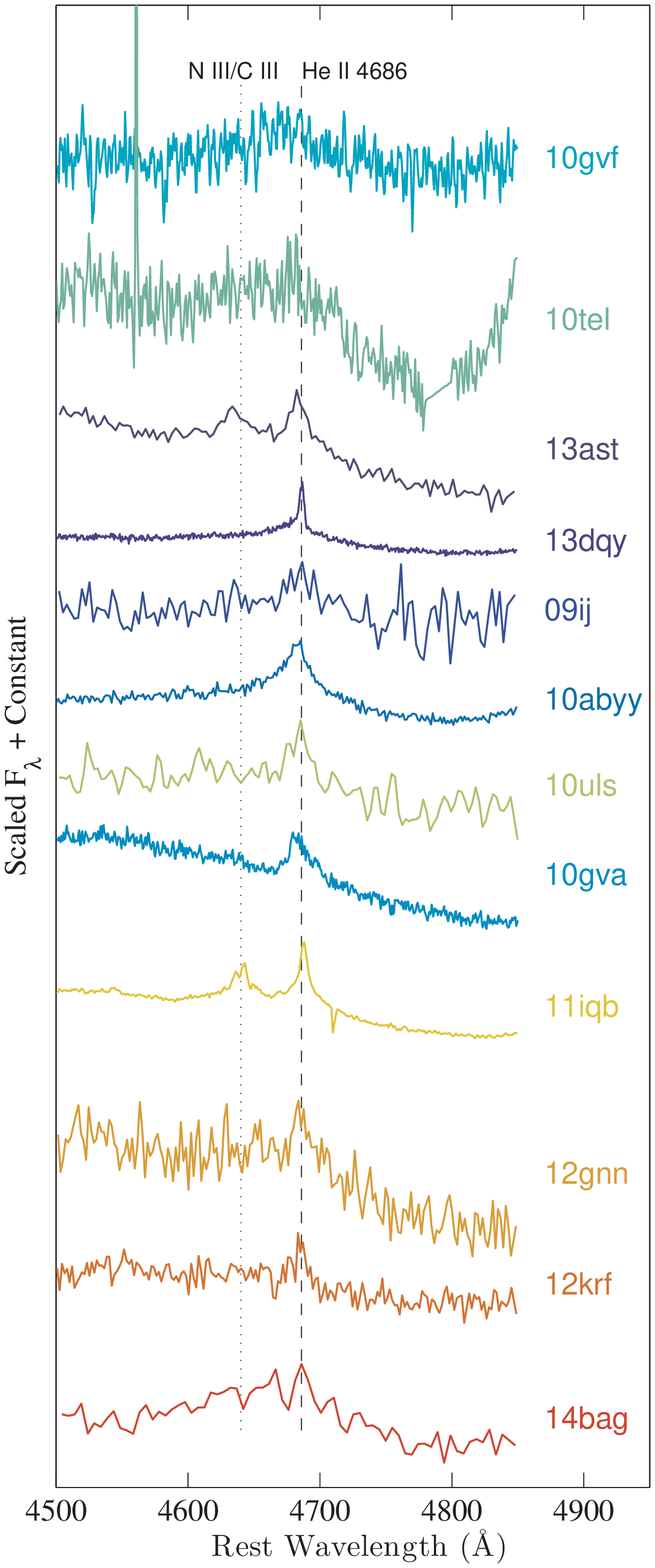}
%
\includegraphics[width=0.3\textwidth]{\pth 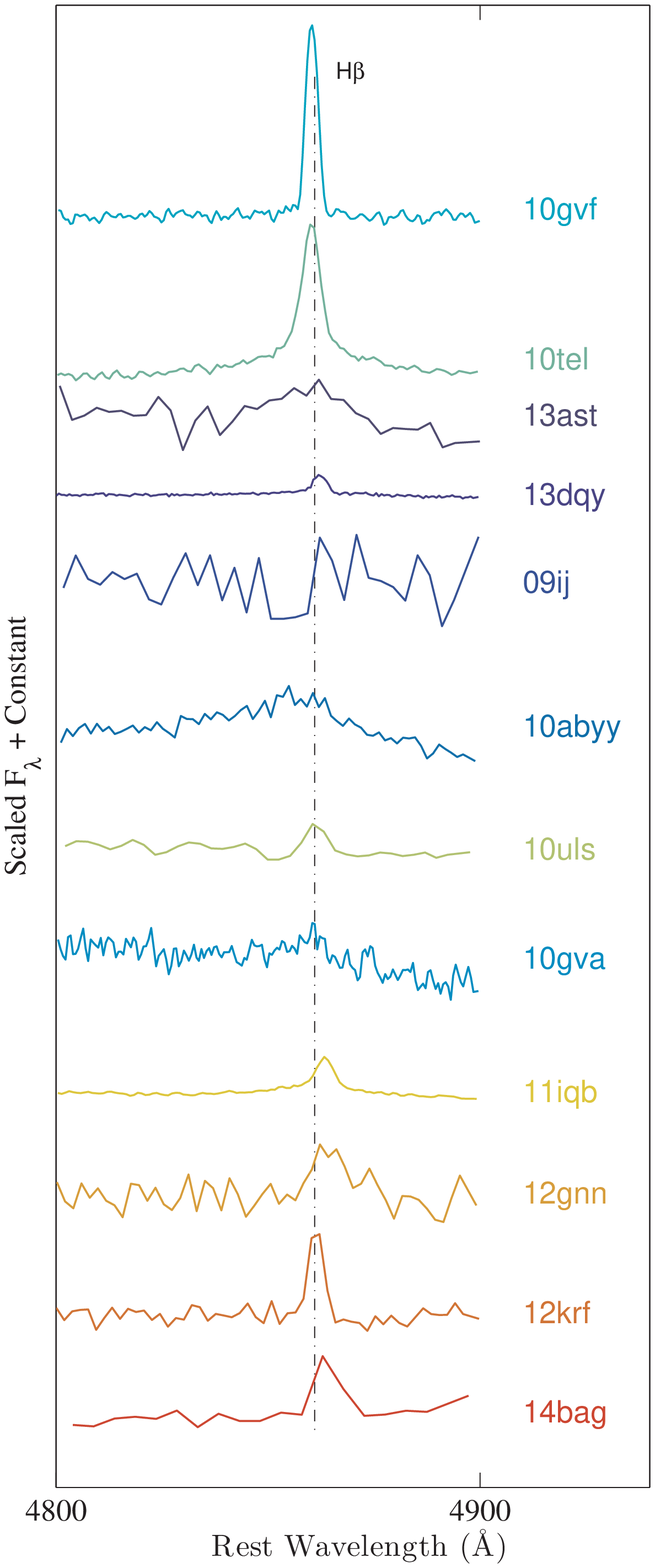}
%
\includegraphics[width=0.3\textwidth]{\pth 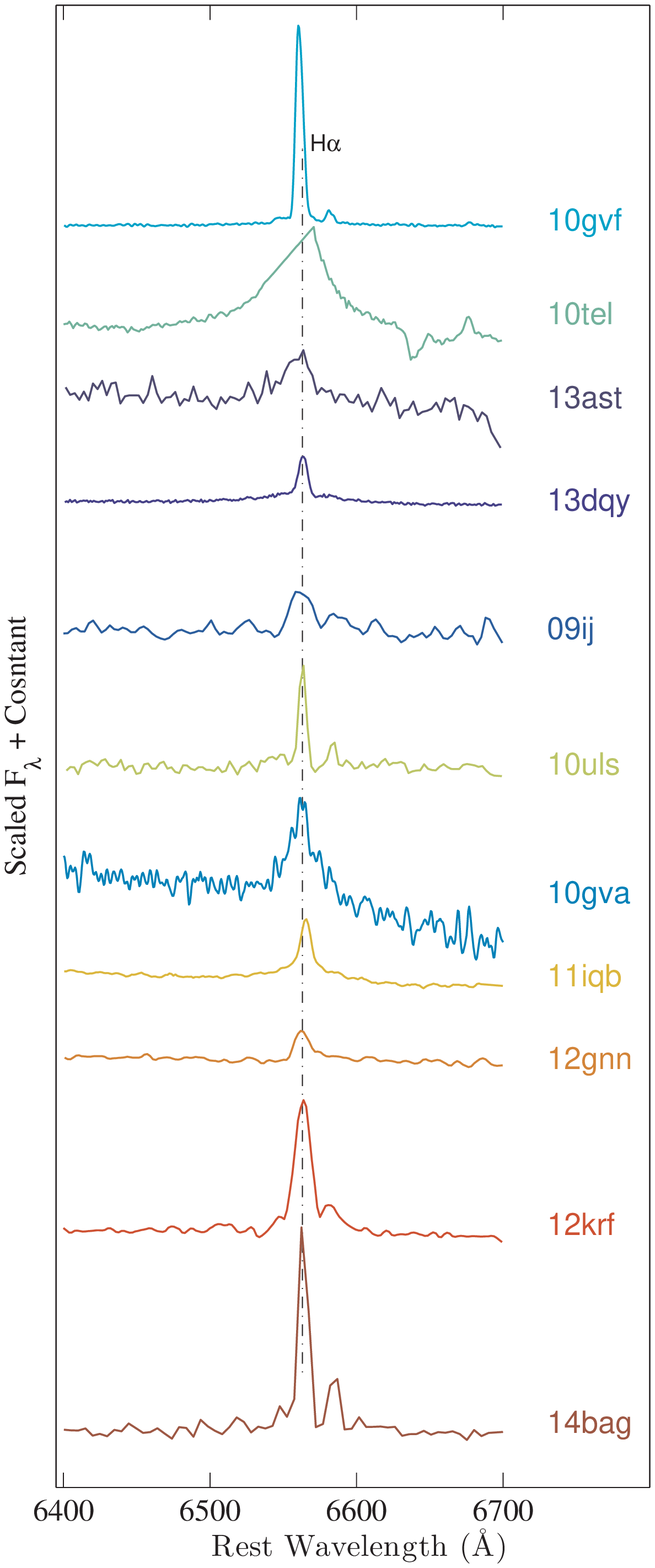}



\caption{Close-up views of the H$\alpha$, H$\beta$, and He~II regions in the first spectra of FI events.}
\label{fig:flashSpecZoomed}
\end{figure*}

\begin{figure*}[h]
\centering
\includegraphics[width=1\textwidth]{\pth 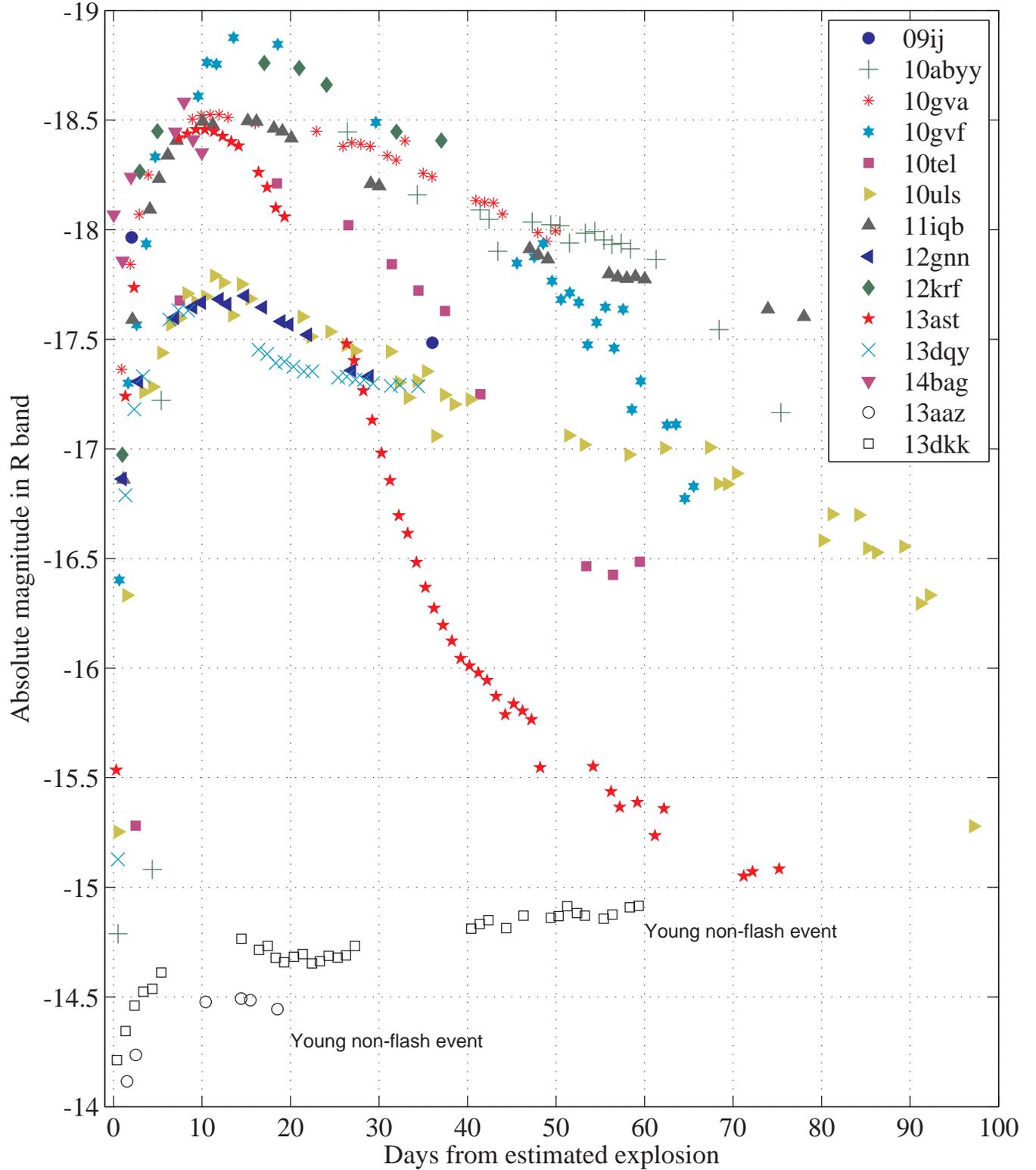}
\caption{$R$-band photometry of the twelve flash-spectroscopy events and two (out of the three) events showing neither FI signatures nor a BF continuum in their spectra $<2$ days after explosion. The light curve of the highly reddened iPTF14ayo is very faint; it is omitted for clarity.  Measurements from the same night have been averaged into a single data point. Error bars are omitted for clarity and are available with the full photometry in the electronic table.}

\label{fig:flashPhot}
\end{figure*}

The distribution by type of the entire sample, the BF SNe, and the FI events is presented in Figure \ref{fig:pie}. Among the 12 FI events, one is of Type IIb, two are SNe~IIn, and the rest are SNe-II.
The spectra of FI events are shown in Figure \ref{fig:flashSpec}, and close-up views of the He~II, H$\alpha$, and H$\beta$ ranges are presented in Figure \ref{fig:flashSpecZoomed}. The corresponding photometry is presented in Figure \ref{fig:flashPhot}.

\begin{figure*}[h]
\centering
\includegraphics[width=1\textwidth]{\pth 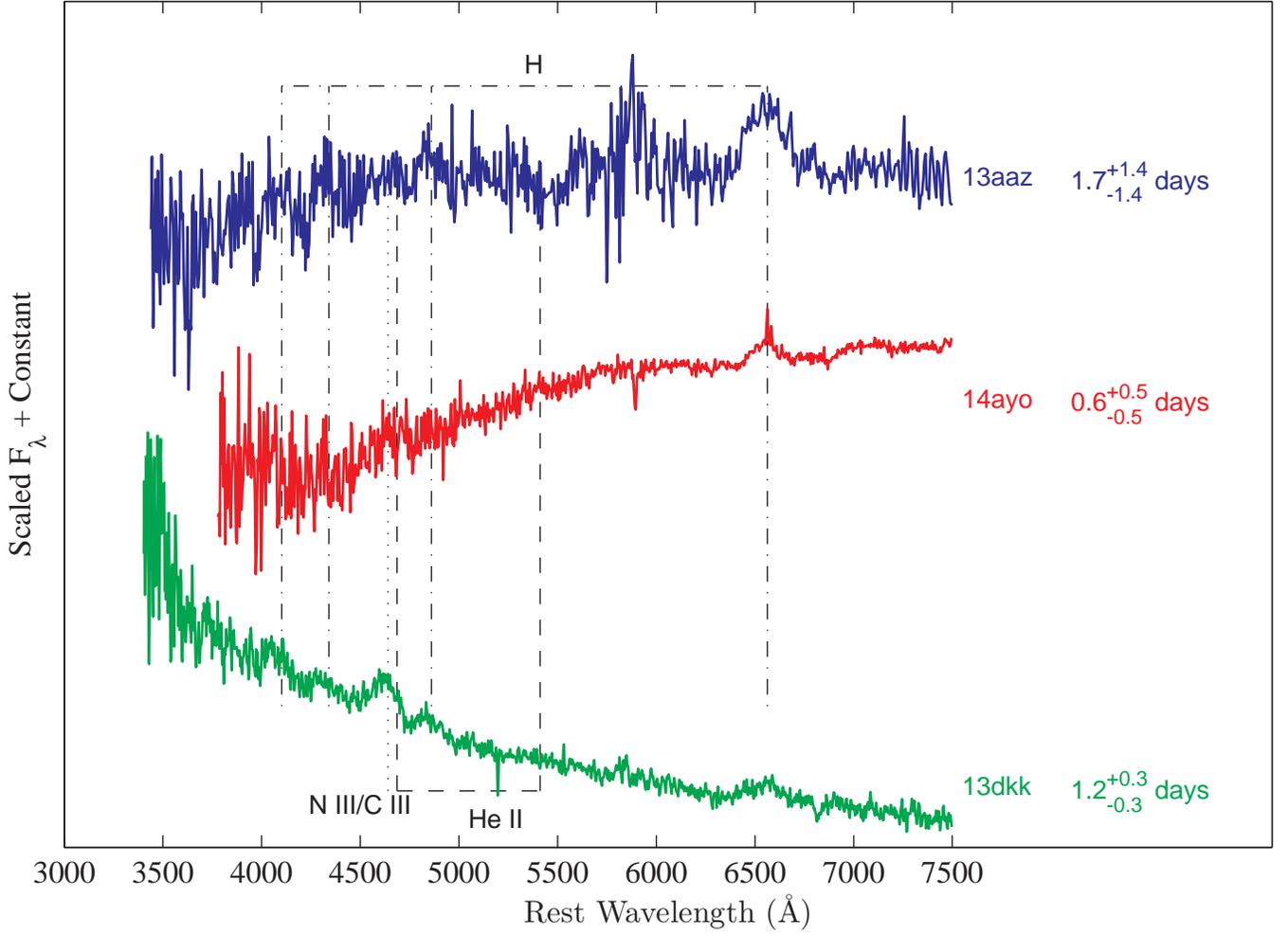}
\caption{The first spectra of SNe that are $<2$ days old and were not classified as FI or BF. At right: the SN age, with respect to the estimated time of explosion (see Appendix \ref{appen:sampleConstruction} for details).}
\label{fig:nonflashSpec}
\end{figure*}

Within two days after the estimated explosion, we found three SNe which have neither FI signatures nor a blue continuum in their first spectra; see Figure \ref{fig:nonflashSpec}. 
iPTF13aaz and iPTF13dkk have already developed broad lines $< 2$ days after explosion. iPTF14ayo is highly reddened, so its true spectroscopic nature is unclear.
The photometry of two of the events can be found in Figure \ref{fig:flashPhot}, while iPTF14ayo, owing to high reddening, has a very faint ($>-12$ mag) light curve that is omitted from the plot. 

Well-observed FI events having a series of spectra taken a few days apart demonstrate that the SN FI signatures disappear over time, making the spectrum featureless, and subsequently P-Cygni lines develop (Figure \ref{fig:flashSpecEvolution}).


\subsection{Fractions}

\begin{figure*}[h]
\centering
\includegraphics[width=1\textwidth]{\pth 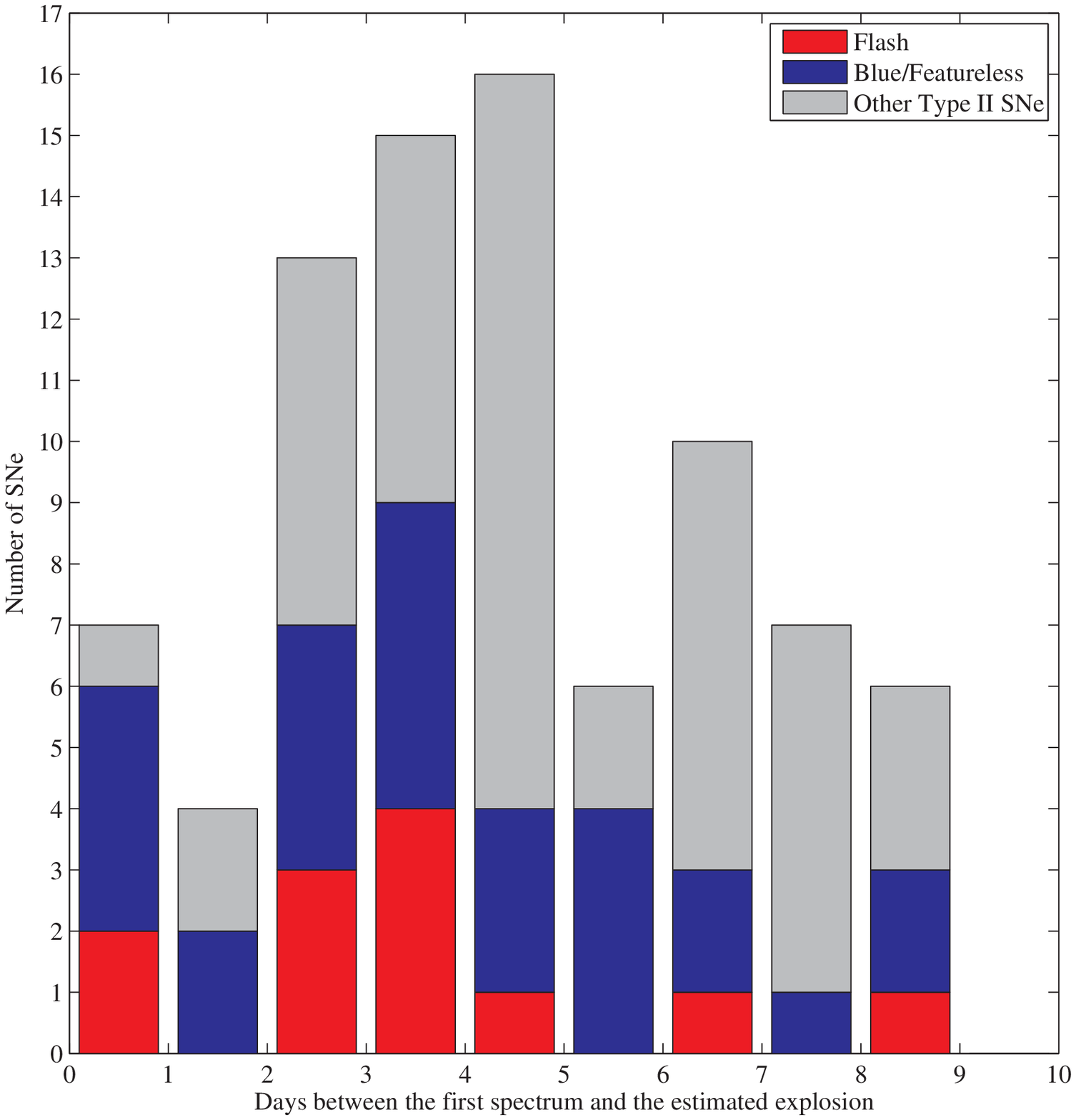}
\caption{Number of SNe~II with respect to the time between the first spectrum and the explosion. FI features appear even at relatively late phases, up to 9 days.}
\label{fig:histLimits}
\end{figure*}

A histogram showing the number of FI and BF events with respect to the age in days after explosion is presented in Figure \ref{fig:histLimits}. Event fractions are given in Table \ref{tab:fractions}.

\begin{table}[h]

\caption{Event Fractions}
\begin{center}
\begin{tabular}{cccc}

Days from explosion & Sample Size & FI & BF\\
\hline \hline
9 & 84 & 14\% & 32\%\\
5 & 55 & 18\% & 33\%\\
2 & 11 & 18\% & 54\%\\
\end{tabular}
\end{center}

\label{tab:fractions}
\end{table}

Assuming a binomial distribution where an event can be classified either as FI or not, the 68\% confidence interval on the fraction of FI events given the available sample is in the range 7--36\%, 13--24\%, and 11--19\% within 2, 5, and 9 days after the explosion, respectively.


\begin{figure*}[h]
\centering
\includegraphics[width=1\textwidth]{\pth 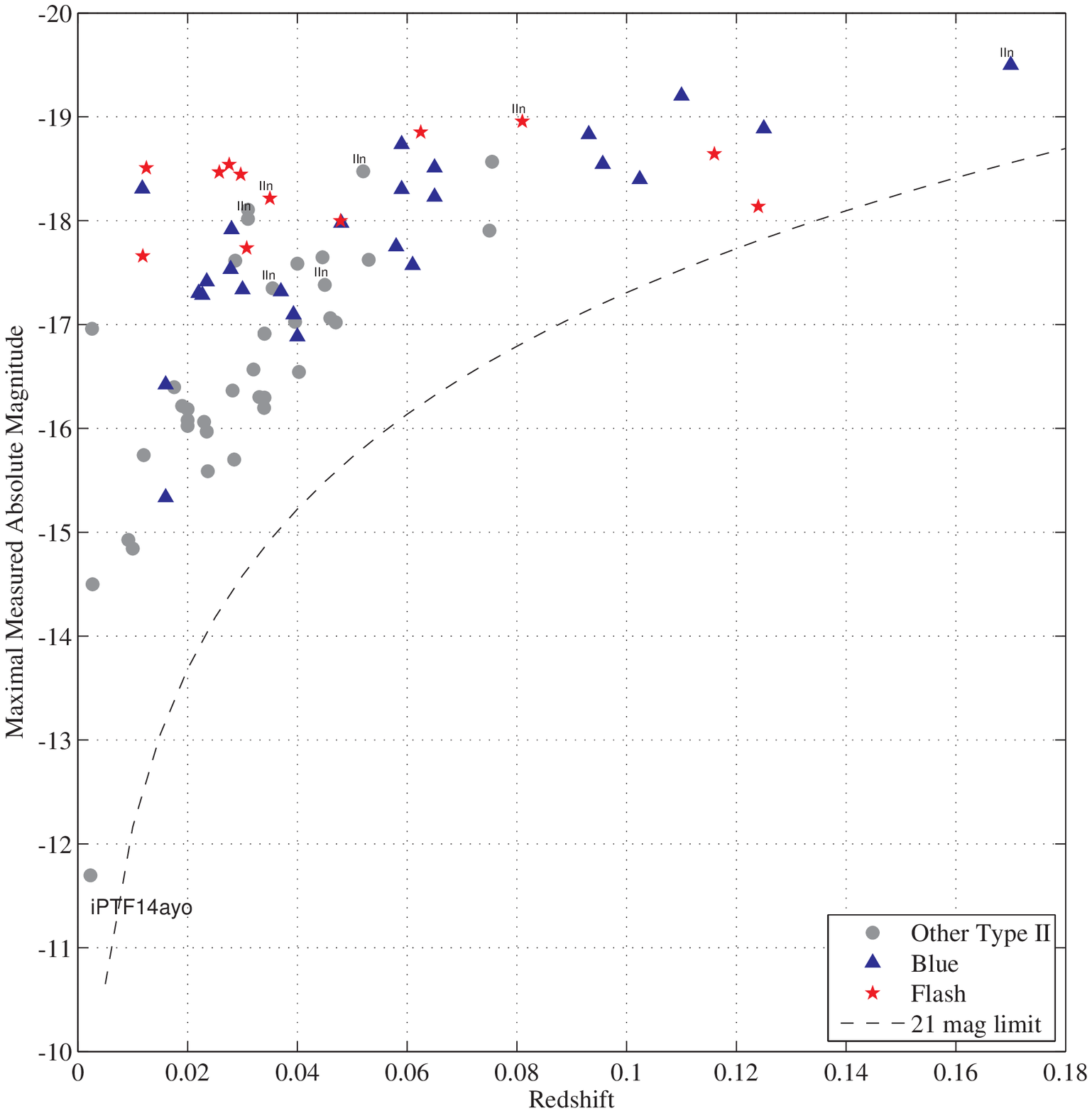}
\caption{The maximal measured absolute magnitude in the $R$ band (9 events with only early $g$-band photometry are omitted from this plot; two of these are BF) vs. the redshift. The dashed line corresponds to our average detection limit of $R=21$\,mag. iPTF14ayo is very faint, probably because of high extinction. FI and BF events dominate the high-luminosity end.}
\label{fig:absMagVsRedShift}
\end{figure*}

\section{Discussion}
\label{sec:discussion}


The FI fractions we found are significant ($\sim 15$\%), yet these are clearly lower limits on the true fractions. The SNe which initially show FI signatures develop BF spectra at some later point (Figure \ref{fig:flashSpecEvolution}). It is quite possible that the SNe showing a blue continuum in their first spectrum would have exhibited FI signatures at an earlier phase. Therefore, if we had earlier spectra of the entire sample, we would have observed a higher fraction of FI events. It is also possible that FI events were not detected owing to the low signal-to-noise ratio in some of the spectra. In addition, there is possible contamination by slowly rising old events which appear to be younger, and satisfy our cut of the pre-explosion limit being only 0.5\,mag deeper than the detection (see Appendix \ref{appen:sampleConstruction} for details). Future samples of events with systematic early-time spectroscopy should provide a more accurate estimate of the fraction of SN progenitors embedded in dense CSM. Our work demonstrates that this is not a rare occurrence.

An interesting result can be seen by inspecting Figure \ref{fig:absMagVsRedShift}, which shows the natural connection between the redshift and the SN peak luminosity\footnote{Luminosity as measured, not necessarily the peak of the true SN light curve.}, as expected in flux-limited surveys like PTF. SNe with FI signatures or a BF spectrum are more luminous on average than the other SNe in our sample. This is true even when we exclude the three SNe~IIn which are classified as FI or BF. Using the Kolmogorov-Smirnov test, we reject the possibility that the peak-magnitude distribution of the FI and BF SNe is the same as the rest of the events, with a confidence level higher than 99.9\%. Since SNe showing FI signatures or a BF continuum in their spectra are typically hotter (as hydrogen remains ionized and broad lines do not appear), we can conclude that more-luminous SNe have higher temperatures at comparable ages than less-luminous ones. It is unlikely that the high peak luminosity is caused by the interaction with the CSM (excluding SNe~IIn), since the emission lines usually do not last for more than 10 days, while the events typically maintain their high luminosity for $>20$ days.

To estimate the fraction of FI and BF events in a volume-limited survey, we inspected all events at $z<0.02$, where our survey is sensitive to SNe with a peak absolute magnitude $R<-14$ (Figure \ref{fig:absMagVsRedShift}). Within the statistical uncertainty, we obtain the same fractions as for the entire sample.
Our accuracy is limited by the small number of events at $z<0.02$ (just 14). A larger sample may show lower fractions of FI and BF events in a volume-limited survey, since flux-limited data miss mostly the low-luminosity events that are usually neither FI nor BF.



Type II-P events are the most common among CC~SNe \citep{2009MNRAS.395.1409S}, and constitute almost 30\% of CC~SNe in an ideal magnitude-limited sample \citep[][their Figure 11]{2011MNRAS.412.1441L}. Typically SNe~II-P display a plateau in their light curve for $\sim 100$ days \citep{2012ApJ...756L..30A}, with an absolute magnitude of $M_R\approx -16$. However, Figure \ref{fig:flashPhot} shows that none of the FI events has such a light curve. This suggests that either typical SN~II-P progenitors have little CSM and do not produce FI features, or  such features might be shorter lived for SNe~II-P and have been missed in our dataset. This is compatible with the finding that the spectrum of the Type II-P SN 2006bp, which is among the earliest yet obtained for objects of this class \citep{2007ApJ...666.1093Q, 2011ApJ...736..159G}, did show FI features and would have been picked up by our selection criteria (Table \ref{tab-spec}).
Indeed, two of the SNe younger than 2 days which are not FI or BF have light curves that are similar to those of faint SNe~II-P, and clearly differ from those of the FI events.




With early-time spectra that resemble those of iPTF13ast, the FI events show clear evidence for the existence of CSM around the progenitor. Assuming this material was expanding at a velocity of $\sim 100$\,km\,s$^{-1}$ \citep{2014A&A...572L..11G}, and that the emission lines disappear after being swept away by the expanding ejecta moving typically at $10^4$\,km\,s$^{-1}$, and given that all of the FI events are constrained to be within 10 days after the explosion, we can conclude that such a CSM was ejected from the star $\sim 1000$ days prior the explosion.
SN spectra within one day after explosion may show additional emission lines (such as N~IV; \citealt{2014Natur.509..471G}) which disappear in only 1--2 days, so they are crucial in order to obtain the complete chemical composition of the CSM. Such early spectra also probe the nearby CSM, which was ejected from the progenitor just prior to the explosion. Subsequent early spectra would allow us to determine the temperature evolution of the SN. All of these reasons strongly motivate observing SNe as soon as possible after explosion.

\section{Summary}
\label{sec:summary}
Motivated by the discovery of iPTF13ast showing prominent high-ionization emission lines in its early-time spectra, we searched the PTF and iPTF databases from 2009 until the end of 2014 for similar events. We found that FI signatures typically occur in Type II SNe, so we constructed a sample consisting of 84 Type II SNe whose first spectra are constrained to be within 10 days after explosion. 

We classified the events in our sample according to their first spectra as FI, BF, or neither. We found that 14\% of the SNe in our sample have FI signatures, whereas within five and two days after explosion the fraction is 18\%.
The actual fraction of FI events is likely higher, since FI SNe evolve into BF events as time progresses, and BF events develop normal SN features later on. Earlier spectra of a similar sample of events are expected to yield a higher fraction of FI events.

We obtained an interesting result regarding the connection between the SN temperature and its luminosity.
In Figure \ref{fig:absMagVsRedShift} we see that at relatively high redshifts we have only BF and FI events. Those distant events are more luminous than average SNe~II, as we expect from a flux-limited survey. However, we see in addition that 19 out of 21 events brighter than $M_R=-18.2$ mag are classified as BF or FI events. 
Since the lack of hydrogen lines in BF events and the existence of the He~II emission line in FI events are an indication of high temperatures, and there are no interaction signatures $>10$ days after explosion, we conclude that more-luminous SNe maintain higher temperatures during their early-time evolution.


Using flash spectroscopy it should be possible to probe the material ejected from the progenitor $\sim 1000$ days prior its explosion, reveal its elemental composition, track the SN early-time temperature evolution, and find the progenitor mass-loss rate shortly before the explosion. As our study shows, such events are not rare; thus, application of this method to future samples is a promising prospect.

\acknowledgements

We would like to acknowledge A.~Miller, D.~Poznanski, J.~Bloom,
K.~Clubb, D.~Murray, S.~Tang, S.~Geier, V.~Van~Eylen, and J.~Parrent
for help with the observations.  We are grateful to the staffs at the
various observatories and telescopes with which we collected data (Lick,
Palomar, Keck, Gemini, NOT, Apache Point, Kitt Peak, HET) for their
excellent assistance.  An anonymous referee careful read our
manuscript and provided valuable comments.

A.G.Y. is supported by the EU/FP7 via ERC grant no. 307260, the Quantum Universe I-Core program by the Israeli Committee for planning and budgeting, and the ISF, Minerva and ISF grants, WIS-UK ``making connections,'' and the Kimmel and ARCHES awards. J.M.S. is supported by a National Science Foundation (NSF) Astronomy and Astrophysics Postdoctoral Fellowship under award AST-1302771. 
A.V.F's research was made possible by NSF grant AST-1211916, the TABASGO Foundation, and the Christopher R. Redlich Fund.
M.S. acknowledges support from the Royal Society and EU/FP7-ERC grant no [615929].
LANL participation in iPTF is supported by the US Department of Energy
as part of the Laboratory Directed Research and Development program.
A portion of this work was carried out at the Jet Propulsion Laboratory under a Research and Technology Development Grant, under contract with the National Aeronautics and Space Administration. US Government Support Acknowledged.

Research at Lick
Observatory is partially supported by a generous gift from Google. 
Some of the data presented herein were obtained at the W. M. Keck
Observatory, which is operated as a scientific partnership among the
California Institute of Technology, the University of California, and
NASA; the observatory was made possible by the generous financial
support of the W. M. Keck Foundation.
Based in part on observations obtained at the Gemini Observatory under program GN-2010B-Q-13, which is operated by the Association of Universities for Research in Astronomy, Inc., under a cooperative agreement with the NSF on behalf of the Gemini partnership: the NSF (United States), the National Research Council (Canada), CONICYT (Chile), the Australian Research Council (Australia), Minist\'{e}rio da Ciencia, Tecnologia e Inovacao (Brazil), and Ministerio de Ciencia, Tecnolog\'{i}a e Innovaci\'{o}n Productiva (Argentina).

\appendix
\section{Sample Construction}
\label{appen:sampleConstruction}
We started by looking for CC~SNe which had a spectrum obtained within 6 days after PTF discovery\footnote{Note that the discovery date can be a long time after the actual explosion.} and found 332 events.
All of the spectra were then classified as either blue/featureless (BF) events, ``flash-ionized" (FI) events, or neither.

All 332 events were processed through our photometric pipeline in order to obtain a better estimate of the SN explosion time, and in many cases we found detections earlier than the PTF discovery.


In order to constrain the SN explosion date and the time of the pre-explosion limit we did the following:
\begin{enumerate}

\item
By default, we chose the date of the latest nondetection where the limiting magnitude (3 times the flux uncertainty, converted to magnitude) is fainter by at least 0.5\,mag from the magnitude of the first detection.
We then adopt an explosion date which is the mean date between the last nondetection and the first detection.

\item
In the few cases where the spectrum was taken less than 10 days after the first detection, the pre-explosion limit was at least 1 day before the detection, and we had well-sampled photometry along the rise ($\geq 4$ nights of data before the peak). We calculated a parabolic fit to the data (all the data points along the rise), and chose the date when the flux was 0 as the explosion date. There were 19 such events in total; 3 of them are classified as FI and 5 as BF.

\item
In the case of PTF10tel (SN 2010mc), which had a precursor \citep{2013Natur.494...65O}, we adopt as the SN explosion the mean between the date of the last precursor measurement and the date of the first measurement of the SN (main peak).
\end{enumerate}


All twelve events showing FI signatures in their first spectrum were younger than 10 days from the time of the pre-explosion limit, and are Type II SNe. We exclude a peculiar object of uncertain nature that possibly shows FI signatures and is the subject of a forthcoming publication (Kasliwal et al., in prep.). Hence, we omitted from further analysis all the events whose first spectra are not constrained to be within 10 days from explosion, and we focus on SNe~II.  

For one event we were unable to obtain reliable photometry in order to estimate its explosion time. This SN is at least 3 days old and is neither FI nor BF.




In order to check if we missed any additional SNe having a spectrum obtained $<10$ days after explosion, we looked at all the CC~SNe whose first spectrum was obtained within 10 days after PTF discovery. There are $>100$ additional events, but we found only 5 SNe~II that match our selection criteria, and we added them to our sample.

We ended up with 84 SNe~II in our sample, which includes all CC~SNe whose first spectrum is constrained to have been obtained within 10 days after explosion. 

\begin{figure*}[h]
\centering
\includegraphics[width=1\textwidth]{\pth 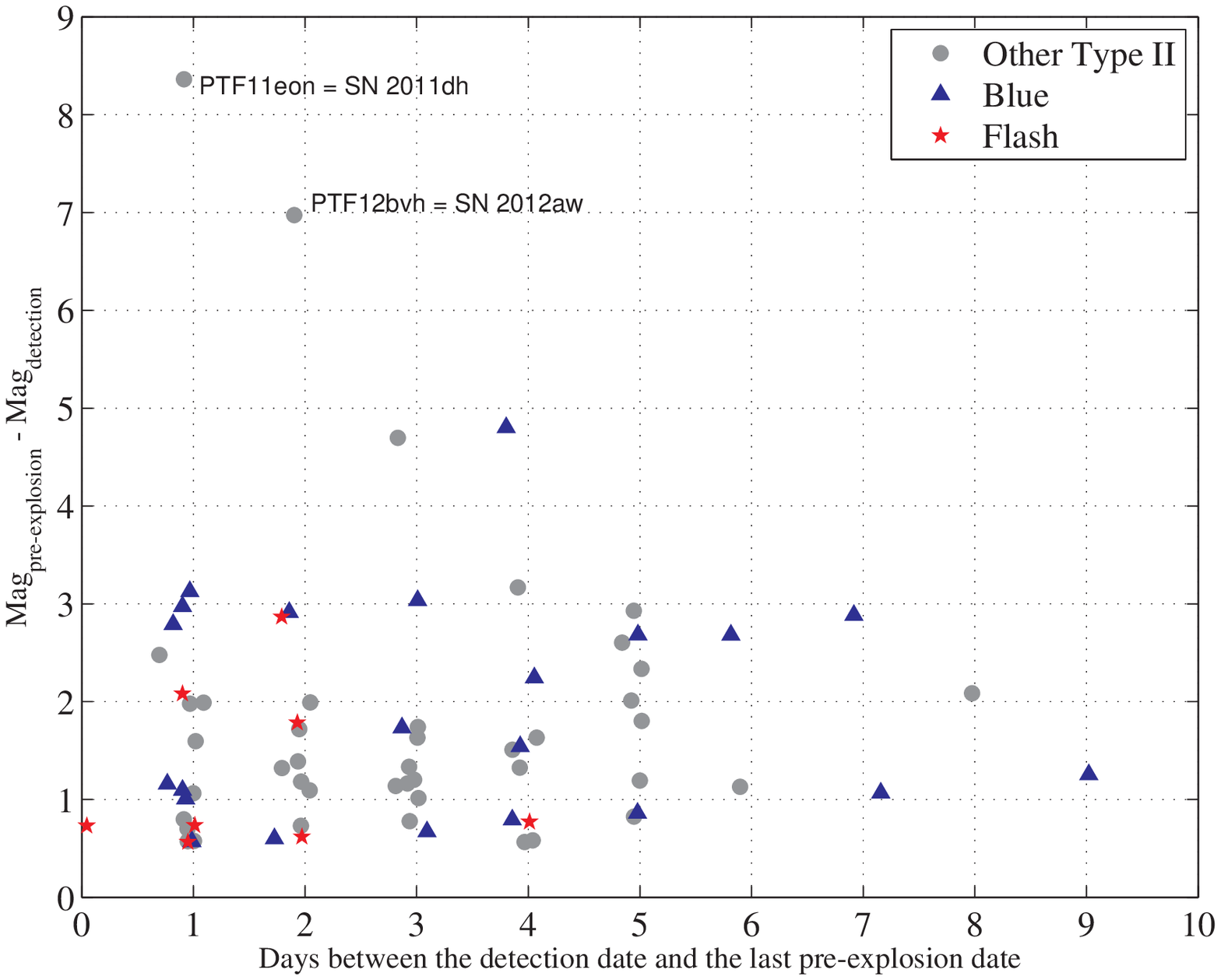}
\caption{The magnitude difference between the pre-explosion limit and the first detection vs. the difference in days between the first detection and the pre-explosion (last nondetection) limit. All of the events with explosion time estimated by fitting the rise time (rather than using a recent pre-explosion limit) are omitted from this plot. SN 2011dh and SN 2012aw are nearby SNe with very deep nondetection limits, hence the large magnitude difference in a short time.}
\label{fig:magDiff}
\end{figure*}

Figure \ref{fig:magDiff} shows the magnitude difference between the first detection magnitude and the limiting magnitude in the pre-explosion limit, versus the time difference between the two. In order to estimate an upper limit on the SN age, we required the limiting magnitude in the pre-explosion limit to be fainter by 0.5\,mag with respect to the first detection. The resulting estimates are consistent with measurements based on well-sampled photometric light curves for all FI events (except PTF09ij, which does not have detailed photometry, and PTF10tel, which had a precursor). However, SNe may rise by $\sim 3$\,mag within 10--20 days after the explosion, so our sample may have some contamination by old, slowly rising events that appear to be younger. This may be seen as the apparent excess of gray points around $\sim 1$\,mag difference, compared to FI/BF events' domination at values of 2--3\,mag. This contamination will result in estimating a lower fraction of FI events than the true one, making our reported fraction a lower limit.

\begin{figure*}[h]
\centering
\includegraphics[width=1\textwidth]{\pth 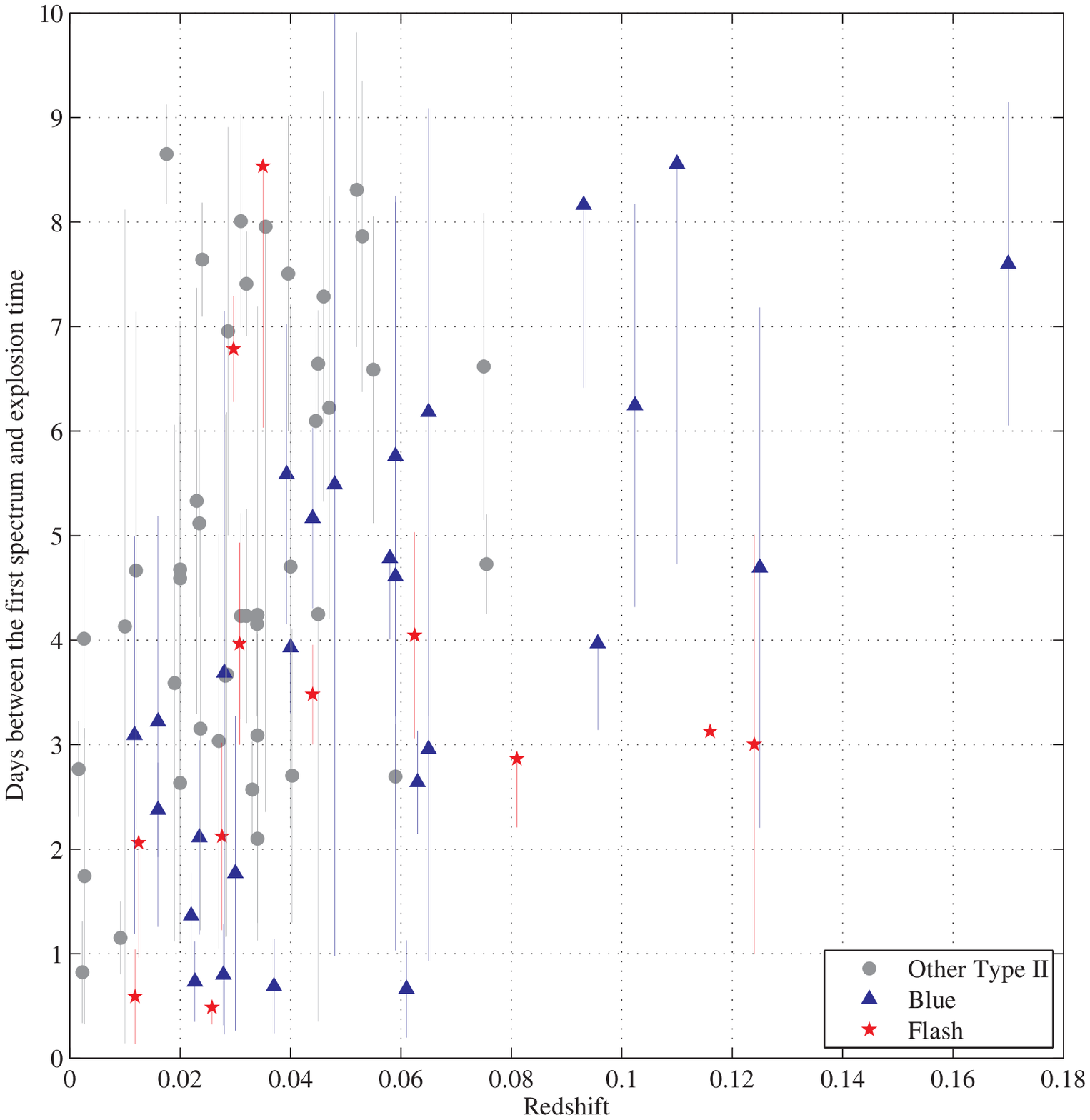}
\caption{Time between the first spectrum and the estimated explosion time vs. the redshift. The error bars are between the times of the first detection and the pre-explosion limit (which is the same as the estimated explosion time for the SNe with fitted rise time). Note the lack of distant SNe that are not classified as BF or FI events.}
\label{fig:redShift}
\end{figure*}

 
For completeness, we present in Figure \ref{fig:redShift} the estimated age of SNe~II in our sample versus their redshift. The error bars span the time between the first detection on the lower end and the pre-explosion limit on the other. Note that for some cases the time of the pre-explosion limit is the same as the estimated explosion time. 

\bibliographystyle{apj}
\bibliography{flashSpecBib}



\end{document}